\documentclass[english]{article}
\usepackage[T1]{fontenc}
\usepackage[latin9]{inputenc}
\usepackage{babel}
\begin{document}
Stretched Magnetic Moments

L. Zamick and Y.Y. Sharon

Department of Physics and Astronomy, Rutgers University, Piscataway,
New Jersey 08854 U.S.A.

We here wish to complie some simple expressions for nuclear magnetic
moments.The results are not new but it may be useful to have things
all in one place.

To make things concrete we use as an example a paper by G. Ile et
al. {[}1{]} . In that work they measuredthe g factor of the lowest
7$^{-}$state of $^{126}$Sn. They note that the leading configuration
is {[}h$_{11/2}$ d$_{3/2}${]}J=7 and they correctly give the value
of the g factor of this configuration as -0.11. 

Note that the above configuration is stretched. That is to sayJ=7
is the largest possible value one can have .For a stretched configuration
$ $ the magnetic moments that add . We now give more details of this
not unknown result.

By convention the magnetic moment of a system of particles with total
angular momentum J is defined as the expectation value of the z component
of the magnetic moment operator in a state with M=J $ $. For a system
of two nucleons with configuration {[}j$_{1,}$j$_{2}${]} with J=j$_{1}$+j$_{2}$
(stretched case) the wave function factorizes.

$\Psi$(j$_{1}$,j$_{2}$,J,M=J) =$\psi$(j$_{1}$,m=j$_{1}$) $\psi(j_{2}$,m=j$_{2}$)

Note that for each of the 2 wave functions on the right we have m=j
which is just what you want for getting the magnetic moment .From
this it obviously follows that $\mu$(J)= $\mu$(j$_{1}$) +$\mu$(j$_{2}$)
for this stretched case. For the above configuration tthe values fort
the 2 terms are $ $$ $-1.913 (free neutron) and -3/5 (-1.913) $ $
so that the total is 2/5 (-1.913).

More generally for any J value the relation satisfied by the g factor
is a bit more complicated: 

g=(g$_{1}$+g$_{2}$)/2 + (g$_{1}$-g$_{2}$) / (2 J(J+1)) {*} {[}j$_{1}$
(j$_{1}$+1) -j$_{2}$ (j$_{2}$+1){]}

For the above configuration one obtains g=-2/35{*}$\mu(free$neutron)=-0.109.
This is close to the experimental value -0.098(9).

In some cases the formula for g simplifies. If j$_{1}$ is equal to
j$_{2}$ we have

g= (g$_{1}$ +g$_{2}$)/2

We also have this result if g$_{1}$=g$_{2}$ . This has the consequence
that all g factors for idetical nucleons in a single j-shell (e.g.
only neutrons) are the same. We also get this result for a system
with an equal number of protons and neutrons in a single j shell;
also equal number of protons and neutron- holes or neutrons and proton
-holes.

{[}1{]} G. Ilie et al., Phys.Lett. B 687 (2010)305 
\end{document}